\documentclass{aastex62}

\submitjournal{ApJ}

\usepackage{amsmath}

\begin{document}

\title{Principle of Minimum Energy in Magnetic Reconnection in a Self-Organized Critical Model for Solar Flares}

\correspondingauthor{Michael S. Wheatland}
\email{michael.wheatland@sydney.edu.au}
\email{farhangnastaran@gmail.com}
\email{safari@znu.ac.ir}

\author{Nastaran Farhang}
\affil{Department of Physics, Faculty of Science, University of Zanjan \\
 P.O. Box 45195-313, Zanjan, Iran}

\author{Hossein Safari}
\affiliation{Department of Physics, Faculty of Science, University of Zanjan \\
 P.O. Box 45195-313, Zanjan, Iran}

\author{Michael S. Wheatland}
\affiliation{Sydney Institute for Astronomy, School of Physics, The University of Sydney \\
 NSW 2006, Australia}



\begin{abstract}

Solar flares are an abrupt release of magnetic energy in the Sun's atmosphere due to reconnection of the coronal magnetic field. This occurs in response to turbulent flows at the photosphere which twist the coronal field. Similar to earthquakes, solar flares represent the behavior of a complex system, and expectedly their energy distribution follows a power law. We present a statistical model based on the principle of minimum energy in a coronal loop undergoing magnetic reconnection, which is described as an avalanche process. We show the distribution of peaks for the flaring events in this self-organized critical system is scale-free. The obtained power law index 1.84 $ \pm $ 0.02 for the peaks is in good agreement with satellite observations of soft X-ray flares. The principle of minimum energy can be applied for general avalanche models to describe many other phenomena.

\end{abstract}

\keywords{Sun: flares; Sun: magnetic field; Sun: self-organized criticality; Sun: waiting time}

`
\section{Introduction} \label{sec:intro}

Twisting of magnetic footpoints as a result of photospheric motions causes stressing of the coronal magnetic field. This may lead to a series of magnetic reconnections (a flare), in which abrupt release of energy occurs \citep{Gold1960, longcope1996, Parker1988, priest2000, craig2002, mendoza2014}. Observations from various spacecraft and balloon-borne detectors including the \textit{Geostationary Operational Environmental Satellites} (GOES), \textit{Yohkoh}, the \textit{Solar and Heliospheric Observatory} (SOHO), \textit{Hinode}, and the \textit{Solar Dynamics Observatory} (SDO), show that solar flare energies span a wide range from $10^{25}-10^{33}$ ergs \citep{Crosby1993, shimizu1995, Aschw2000b, emslie2012, Aschwanden2014, maehara2015}. The size distribution of solar flares follows a power law \citep{mike2001, Litvinenko2001, aschwbook2011}, where by \textquotedblleft size\textquotedblright~ we mean some measure of the magnitude of flares, e.g. the peak flux in soft X-rays. The power-law index lies in the range 1.5-2.1 \citep{georg1998, nita2002, mendoza2014}, depending on the choice of the measure of size. This indicates the scale-free and stochastic nature of solar flares as a self-organized critical system \citep{Crosby1993,charbon2001, aschwbook2011, aschw2014, Aschwanden2014}. It is shown that solar flares are complex systems due to the self-organized criticality, limited predictability, etc. The network of flares which occur on the solar surface has the features of scale-free and small world networks \citep{gheibi2017}. Therefore, it is expected that the statistics of the solar flare complex system such as the size distribution follows power-law behavior.

Based on Parker's theorem, in a self-organized critical system gradual perturbations lead the system to a critical state in which a sequence of bursts takes place \citep{Parker1988}. Cellular automaton avalanche models are applied to study the quasi-static evolution of the coronal magnetic field and investigate whether the flaring events trigger one another or are a completely uncorrelated processes \citep{bak1987, LH1991, LH1993, Wheatland1996, Vassiliadis1998, isliker1998, boffeta1999, isliker2000, charbon2001, hughes2003, barpi2007, strugarek2014}. The cellular automaton method allows us to study the complex behavior of avalanches by dividing them into discrete elements \citep{LH1991,zirker1993,robinson1994,charbon2001,buchlin2003,morales2008a,morales2008b,stru2014,mendoza2014}. A paradigm of an avalanche process is the sandpile model, in which a field defined on a set of nodes is being driven slowly using random numbers \citep{bak1987}. The stability of each node is checked against a criterion after each driving event. In case of an instability at a node, the field is redistributed (following some redistribution rules) to locally relax the system. All redistribution rules in the existing avalanche models are ad hoc, and various models are possible. The principle of minimum energy for natural systems can give a definite model.

Applying the cellular automaton approach, we investigate the complex evolution of the magnetic vector potential field within a flaring coronal loop as a self-organized critical system. We introduce a two-dimensional lattice-based model by defining a magnetic vector potential field on a grid of nodes. In general, the discrete magnetic field does not satisfy $ \boldsymbol{\nabla} \cdot \boldsymbol{B}=0, $ so we use the magnetic vector potential instead. The nodal value of the lattice,
$ A_{i,j}, $
represents the average magnetic vector potential field in each cell. Therefore, the magnetic field and the electric current density can be measured by using the nodal values of each cell and its neighbors. The two-dimensional grid represents a cross section of a flux tube, which includes magnetic strands (for a typical coronal loop divided into 52 $ \times $ 52 cells, each cell is the cross section of a magnetic strand within the flux tube). The topological transformation of the magnetic field (or its equivalent magnetic vector potential) due to the turbulent flows and twisting of the magnetic footpoints is applied to the model through a global driving mechanism. In this model, a magnetic reconnection is considered as a short-distance interaction between an unstable node and its nearest neighbors which renews the nodal values of the group of nodes. Different contributions of interactions between cells are studied. Here, among all unstable nodes, redistributions only take place at nodes maximizing the released energy in the context of the principle of minimum energy for the final equilibrium state.

In Section \ref{sec:style}, we briefly introduce the solar flare observational data, and present our new avalanche model based on optimization of released energy. We propose a new set of redistribution rules to describe the magnetic reconnection over a two-dimensional lattice as a cross section of a coronal loop, which minimizes the magnetic energy of the system after each magnetic reconnection (burst). Then, we apply two new methods, based on genetic algorithm and maximum likelihood estimation in the Bayesian framework, to obtain the power-law indexes of both simulated and observational flaring events. In Section \ref{res}, we show that the energy and waiting-time distributions of released magnetic energies are in good agreement with observational data. We conclude in Section \ref{sec:discussion} by summarizing our new model and its results.

\section{methods} \label{sec:style}

\subsection{Solar Flare Data} \label{sec:data}
The \textit{Geostationary Operational Environmental Satellite} (GOES) system, comprising both ground-based instruments and spacecraft, continuously monitors environmental data. These data are commonly used for space weather forecasting and scientific research. The GOES satellites record solar flares in the 0.1-0.8 nm band. Flare events in this band are categorized into five classes: A, B, C, M, and X, based on the peak flux of the events. The GOES event catalogue also records the start and end time of an event, the maximum time (the time when the maximum flux was recorded), the location (longitude and latitude on the solar surface), integrated flux, and the sunspot region number. In this study, we used the peak flux and the start and end times of the registered GOES flare events in the Lockheed Martin Solar and Astrophysics Laboratory (LMSAL) Latest Event Archive from the C, M and X classes. The LMSAL Latest Event Archive is a superset of the GOES events listed by the National Geophysical Data Center (NGDC). Events are obtained using the Solar Soft Interactive Data Language (IDL) routine $ {\rm ssw\_her\_query.pro} $.

\subsection{A Cellular Automaton Avalanche Model for Solar Flares} \label{sec:models}
We introduce a new cellular automaton avalanche model in order to simulate magnetic relaxation (flaring events) in the stressed coronal magnetic field through magnetic reconnection as the magnetic strands within a flux tube interact with each other. From the physical point of view, magnetic potential fields do not possess free energy to liberate. Thus, a non-potential magnetic field with a non-zero electric current is required to explain the energy release in flares. We consider a magnetic flux tube directed along the $ z-$axis with a background field $ \boldsymbol{B}_{0}=B_{0}(r)\hat{\rm{\bold{z}}}. $ The total magnetic field in cylindrical coordinates $ (r,\phi,z) $ is:
\begin{equation}
\label{eq1}
\boldsymbol{B}=\boldsymbol{\nabla} \times (A(r, \phi, t)\hat{\rm{\bold{z}}})+ B_{0}(r)\hat{\rm{\bold{z}}},
\end{equation}
where $ A(r, \phi, t) $ is the magnetic vector potential \citep{stru2014}. This field is in general non-potential, with an electric current density
$ \boldsymbol{J}=\frac{1}{\mu_{0}}\boldsymbol{\nabla}\times \boldsymbol{B}= -\frac{1}{\mu_{0}}\boldsymbol{\nabla}^{2} (A \hat{\rm{\bold{z}}})=-\frac{1}{\mu_{0}}(\boldsymbol{\nabla}_{\perp}^{2} A) \hat{\rm{\bold{z}}}. $
We construct the two-dimensional magnetic vector potential lattice using uniformly distributed random numbers. Application of random numbers increases the degree of stochasticity in the simulations as a basis for solar flares. We consider an open boundary condition at the boundaries. The lattice is slowly excited applying the global driving mechanism by adding a small driving rate
$ \epsilon $
to every cell at each time step, $ t $:
\begin{equation}
\label{eq2}
A_{i,j}^{t+1}=(1+\epsilon)A_{i,j}^{t} \hspace{6mm} \forall (i,j).
\end{equation}
Within $ t $ iterations of driving, the topological degree of twist increases as the azimuthal component of the magnetic field increases by a factor of $ (1+\epsilon)^{t}. $  After each excitation step, the stability is checked over the lattice by introducing an instability threshold
$ Z_{c} $. The threshold is a random number generating from a Gaussian distribution and extracted at each time step. This introduces greater stochasticity. Important parameters in this scheme are the average
$ (\overline{Z_{c}}) $
and width at half-maximum
$ (\sigma) $
of the Gaussian distribution. The instability criterion is defined as:
\begin{equation}
\label{eq3}
\triangle A_{i,j}\equiv A_{i,j} - \frac{1}{4}\sum_{k=1}^{4} A_{k} > Z_{c},
\end{equation}
where the sum runs over nearest neighbors. Among all possible redistribution candidates (unstable nodes) redistribution only takes place at nodes that maximize the released energy. The principle of minimum energy states that during the evolution of a natural system, the maximum possible energy should be released to lead the system to a final equilibrium at the minimum energy. We refer to this model as the optimized model. Thus, if the instability criterion at a node exceeds the threshold, the magnetic reconnection (redistribution) may not occur unless the system releases the maximum amount of energy. This constraint affects the number of peaks and also the inter-flare times (waiting times).
We define the redistribution rules as:
\begin{eqnarray}
\label{eqq5}
A_{i,j}^{n+1}=&&A_{i,j}^{n}- \frac{4}{5} Z_{c}, \vspace{13mm} \nonumber \\
A_{i,j-1}^{n+1}=&&A_{i,j-1}^{n}+ \frac{4}{5}\frac{r_{1}}{x+a}Z_{c}, \vspace{16mm} \nonumber \\
A_{i+1,j}^{n+1}=&&A_{i+1,j}^{n}+ \frac{4}{5}\frac{r_{2}}{x+a}Z_{c}, \vspace{16mm} \nonumber \\
A_{i,j+1}^{n+1}=&&A_{i,j+1}^{n}+ \frac{4}{5} \frac{r_{3}}{x+a}Z_{c}, \vspace{16mm} \nonumber \\
A_{i-1,j}^{n+1}=&&A_{i-1,j}^{n}+ \frac{4}{5} \frac{x}{x+a}Z_{c},
\end{eqnarray}
where
$ r_{1}, r_{2}, r_{3}  $
are uniformly distributed random numbers,
$ a=r_{1}+r_{2}+r_{3}, $
and
$ x $
is the unknown free parameter in the modeling. This parameter is determined to satisfy the optimization constraint. The introduced redistribution rules of Equation (\ref{eqq5}) give:
\begin{eqnarray}
\label{eq66}
\sum A_{i,j}^{n}=\sum A_{i,j}^{n+1}.
\end{eqnarray}
Equation (\ref{eq66}) states that the magnetic vector potential is conserved during each redistribution, which results in conservation of magnetic helicity during reconnections. Generally, it is proved that the magnetic helicity is approximately conserved in dissipative events such as flares \citep{Berger1999}.

As noted above, the vertical component of the non-dimensional electric current density may be written as:
\begin{eqnarray}
\label{eq6}
J_{z}\bigg|_{i,j}
=&& 4A_{i,j}-A_{i+1,j}-A_{i-1,j}-A_{i,j+1}-A_{i,j-1},
\end{eqnarray}
where a centered difference approximation is used for the derivatives. The instability criterion of Equation (\ref{eq3}) can be interpreted as a threshold for the vertical component of the electric current density: $ J_{z}>J_{c} $. The field is redistributed when the current density is locally large. Each redistribution may lead the system to another unstable state. A sequence of redistributions from initial instability somewhere to final stability everywhere is an avalanche, which is identified as a flare. Following an avalanche, the global driving is applied again. This process continues for \textit{t} iterations of driving (time steps).

We consider two methods for calculation of the magnetic energy of the lattice. Previous avalanche models have used an approximate expression for calculation of the lattice energy:
\begin{eqnarray}
\label{eq666}
E=\sum A^{2}.
\end{eqnarray}
In our modeling we use this, as well as the more accurate expression:
\begin{eqnarray}
\label{eq6666}
E=\sum \frac{1}{2} \boldsymbol{A}\cdot\boldsymbol{J}.
\end{eqnarray}
Figure \ref{fig1} explains the role of neighboring cells in the amount of released energy for each method of calculating the energy. In the case of assuming the lattice energy is proportional to $ A^{2} $, the released energy during a redistribution involves the nodal values of 5 cells that are being redistributed. On the other hand, presuming the lattice energy as the dot product of the electric current and the magnetic vector potential field, nodal values of 13 nearest neighbors will affect the released energy based on Equation (\ref{eqq5}).

The liberated energy at each time step is $ \Delta E=E^{t+1}-E^{t} $. The value of $ E^{t} $ is independent of the parameter $ x $ in both methods. Thus, the maximization of the released energy in terms of $ x $ depends only on $ \frac{\partial E^{t+1}}{\partial x} $.

The second derivative test on the released energies leads to two different cases. We discuss the behavior of the magnetic vector potential lattice under these two conditions: (i) the system releases maximum energy at each redistribution, which leads to maximization of the total liberated energy during a flare, (ii) the released energy at each redistribution is an extremum, either minimum or maximum.
\begin{itemize}
\item Case A: We assume $ E = \sum\frac{1}{2} \boldsymbol{A}\cdot\boldsymbol{J}. $ If
$ \frac{\partial^{2} E^{t+1}}{\partial x^{2}} > 0 $,
then
$ E^{t+1} $
has a local minimum at
$ x $.
Therefore, minimization of
$ E^{t+1} $
leads to maximization of
$ \Delta E $,  and vice versa. Mathematical calculations show that for
$ E = \sum\frac{1}{2} \boldsymbol{A}\cdot\boldsymbol{J}, $ the second derivative of $ E^{t+1} $ is always minimum. This causes the maximum exchange of energy between nodes. Although $ \Delta E $ is always maximized for $ E = \sum\frac{1}{2} \boldsymbol{A}\cdot\boldsymbol{J}, $ it should be noted that not all the unstable nodes are allowed to be redistributed. An unstable node is allowed to redistribute only if both $ \frac{\partial^{2} E^{t+1}}{\partial x^{2}} > 0 $ and $ \Delta E <0 $ are satisfied, simultaneously.

\item Case B: We assume $ E =\sum A^{2}. $ In this case, the second derivative of
$ E^{t+1} $ can posses both positive and negative values. The value of $ x $ is determined in a way that the released energy at each local redistribution is an extremum, which can be either a maximum or minimum. In this case, the total liberated energy of all redistributions in one time step is not required to be a minimum or maximum. This is a non-physical approach and our numerical analysis shows that this method is not consistent with the observational data. Therefore, hereafter we omit this case.

\item Case C: We again assume $ E =\sum A^{2}, $ but as a particular situation of case B, the system is allowed only to release a maximum amount of energy. In other words, only nodes that maximize the amount of released energy are allowed to redistribute.
\end{itemize}

\subsection{Application of Genetic Algorithm to Estimate the Power-Law Index} \label{TH:MLE}
Empirical evidence implies the ubiquitous presence of power-law distributions in various fields. This has inspired wide research in physics, geophysics, biology, social sciences, etc., such as studies on self-organized systems \citep{bak1987, newman2005, clauset2009, aschwbook2011, alipour2015}, fractal geometry \citep{mandelbrot1975}, and the scale-free and small-world complex networks \citep{barbasi2003, abe2006, daei2017, gheibi2017}.

The power-law behavior of a quantity $ S $ is defined as $ P(S)\propto S^{- \gamma}, $ where $ P(S) $ is the probability distribution and $ \gamma $ is the power-law index. Practically, only a few empirical distributions obey the ideal power law over all of their values. Therefore, various forms of the power-law function have been developed to provide a better interpretation of the experimental data, like a broken power law, a power law with an exponential cutoff, a curved power law, a truncated power law, thresholded power law, etc. \citep{newman2005, johans2006, aschwa2015}.

Factors like the failure to identify small events, background contamination, and a physical instability threshold lead to deviation of the probability distributions from the ideal power-law behavior at small sizes \citep{aschwa2015}. A thresholded power-law function is often more applicable than a simple power law. The thresholded power-law distribution for a data set with event sizes $ S=s_{1}, s_{2}, ..., s_{M} $ is introduced as:
\begin{eqnarray}
\label{eq9}
P(S)dS= P_{0}(S+S_{0})^{-\gamma}dS,
\end{eqnarray}
where $ S_{0} $ and $ P_{0} $ are the threshold and normalization constant, respectively. To allow the possibility of a finite size effect \citep{aschwa2015} we normalize over a finite range $S_{1}\le S \le S_{2},$ where $ S_{2} $ is the largest event in the data. In that case:
\begin{eqnarray}
\label{eqq101}
P_{0}=(\gamma-1){[(S_{0}+S_{1})^{-\gamma+1} - (S_{0}+S_{2})^{-\gamma+1}]}^{-1}.
\end{eqnarray}
The corresponding cumulative distribution (the probability of obtaining an event larger than $ S,$ but less than $ S_{2} $) is
\begin{eqnarray}
\label{eqq102}
C(S)=&&\int_{S}^{S_{2}} P(S)dS \nonumber \\
=&& \frac{(S_{0}+S_{2})^{-\gamma+1}-(S_{0}+S)^{-\gamma+1}}{(S_{0}+S_{2})^{-\gamma+1}-(S_{0}+S_{1})^{-\gamma+1}}.
\end{eqnarray}
We call Equation (\ref{eq9}) the probability distribution function (PDF) and Equation (\ref{eqq102}) the cumulative distribution function (CDF).

We apply the genetic algorithm \citep{mitchell1996, haupt1998, canto2009, kramer2017} to minimize the $ \chi^{2}$ function:
\begin{eqnarray}
\label{eqq10}
\chi^{2}=\frac{1}{n_{\text{bin}}-3}\sum_{i=1}^{n_{\text{bin}}}\frac{[P_{\text{fit}}(S_{i})-P_{\text{d}}(S_{i})]^{2}}{\sigma_{\text{d}}^{2}},
\end{eqnarray}
where $ n_{\text{bin}} $ is the number of bins, $ P_{\text{fit}}$ is the theoretical PDF (Equation (\ref{eq9})), $ P_{\text{d}} $ is the observational data and/or simulated distribution of each bin, respectively, and $ \sigma_{\text{d}} $ is the expected uncertainty of the observational and/or simulated data size distribution. The uncertainty of the power-law index is $ \sigma_{\gamma}=\gamma / \sqrt{n} $ \citep{aschw2011a}. We minimize the chi-square function with regard to the $ S_{0} $ and $ \gamma $ parameters in order to produce the best thresholded fit, using the genetic algorithm. We also present the cumulative distribution for the thresholded power-law function given by Equation (\ref{eqq102}), but we do not fit data to this model.

\subsection{Maximum Likelihood Estimation in the Bayesian Framework}\label{MLE}
An alternative approach to least squares fitting is maximum likelihood estimation in the Bayesian framework \citep{bai1993, Goldstein2004, newman2005, bauke2007, clauset2009, Giles2011}. Here, we present a new method in the Bayesian framework to infer the power-law index, and the lower threshold. This approach does not face the problem of the arbitrary choice of bins and uses all of the information in the data.

For a data set $ T $ with event sizes $S=s_{1}, s_{2}, ..., s_{M} $ with $S_{1}\le S_{i} \le S_{2},$ and a threshold $ S_{0} $ at small sizes, the normalized distribution is given by Equations \ref{eq9} and \ref{eqq101}. The likelihood function for the data set is:
\begin{eqnarray}
\label{eq12}
P(T\mid S_{0}, \gamma, I)= \frac{(\gamma -1)^{M}}{[(S_{0}+S_{1})^{-\gamma+1}-(S_{0}+S_{2})^{-\gamma+1} ]^{M}} \left(\prod_{i=1}^{M}(S_{0}+s_{i})\right)^{-\gamma}(ds)^{M},
\end{eqnarray}
where $ I $ and $ s_{i} $ are the background information and the event size, respectively. The lower and upper limits of the distributions for the power-law index and the threshold are adopted according to our background knowledge of the data set.

Using Bayes's theorem and a choice of a uniform prior for $ \gamma $ and $ S_{0}, $ the likelihood function of Equation (\ref{eq12}) is converted to the posterior distribution:
\begin{eqnarray}
\label{eq13}
P(S_{0},\gamma \mid T, I)\propto \frac{(\gamma-1)^{M}}{(\gamma_{2}-\gamma_{1})(S_{02}-S_{01})}\frac{\left(\prod_{i=1}^{M}(S_{0}+s_{i})\right)^{-\gamma}}{[(S_{0}+S_{2})^{-\gamma+1}-(S_{0}+S_{1})^{-\gamma+1}]^{M}}.
\end{eqnarray}
Here $ \gamma_{1} $ and $ \gamma_{2} $ denote the range of the uniform prior for $ \gamma $, and $ S_{01} $ and $ S_{02} $ denote the range of the uniform prior for $ S_{0} $. The power-law index can be inferred by integrating Equation (\ref{eq13}) over the undesired parameter $ S_{0} $ as:
\begin{eqnarray}
\label{eq14}
P(\gamma \mid T, I)=\int_{S_{01}}^{S_{02}}P(S_{0},\gamma \mid T, I)dS_{0}.
\end{eqnarray}
The lower threshold can also be extracted by marginalizing the power-law index from the posterior distribution as
\begin{eqnarray}
\label{eq15}
P(S_{0} \mid T, I)=\int_{\gamma_{1}}^{\gamma_{2}}P(S_{0},\gamma \mid T, I)d\gamma.
\end{eqnarray}
The best estimation (mean value) and the uncertainties of the model parameters ($ \gamma $ and $ S_{0} $) are obtained by computing the statistical moments of Equations (\ref{eq14}) and (\ref{eq15}).

\section{Results}\label{res}

\subsection{Model Properties} We present a cellular automaton avalanche model which maximizes the released energy of a flaring coronal loop, in the context of the minimum energy principle for a closed system. In order to simulate the quasi-static evolution of the solar magnetic field, we use a two-dimensional lattice, assumed to represent the magnetic vector potential $ \boldsymbol{A} $ in a cross section of a magnetic flux tube. Uniformly distributed random numbers were generated and used as nodal values of the initial states of the 52 $\times$ 52 lattice. We used the open boundary condition at boundaries to achieve the stationary state in the system. Therefore, the self-organized critical avalanches appear as fluctuations on an approximate stationary state (Figure \ref{fig2}). The full details of the model are given in the Methods section. The magnetic vector potential lattice was subjected to a global driving with a rate
$ \epsilon $,
varied at each excitation step \citep{aschw2010}. Adopted values for $ \epsilon $ were randomly selected in the range of ($ 10^{-7}, 10^{-5} $) from a uniform distribution. The instability threshold,
$ Z_{c} $,
was selected from a Gaussian distribution (with an average of $\overline{Z_{c}}$ and width at half-maximum $\sigma$) renewed at each redistribution step. Application of a small excitement at each time step increases the lattice energy and magnetic helicity of the system. The open boundary condition implies loss of energy at the boundaries. Thus, the system reaches to a stationary state. The magnetic Reynolds number ($ R_{\rm{m}} $) of the solar coronal plasma is large ($ R_{\rm{m}}\approx 10^{8} $) so ohmic dissipation produces only a relatively small rate of dissipation. The open boundary condition is needed to produce a steady state.

Different options for the mean value
$ \overline{Z_{c}} $
and width at half-maximum
$ \sigma $
were studied. It was empirically found that results of case A for a set of simulation with Gaussian parameters of $ \overline{Z_{c}}=1 $, $ \sigma=0.01 $ are compatible with observations.

\subsection{Statistics of Bursts} We calculated the lattice energy at each time step, applying both an accurate method $ E=\sum\frac{1}{2} \boldsymbol{A}\cdot\boldsymbol{J} $ and an approximate method $ E=\sum A^{2}. $ The statistics of the released energies due to reconnections (bursts) is studied for different cases: the case A, which always maximizes released energies using an accurate method, and the case C in which the energy is computed through an approximate method. In the case C only nodes that maximize the released energy are allowed to redistribute. Our results show that confining the system to release some specific amount of energy affects the number of flaring events. In previous models (for example \cite{stru2014}) five degrees of stochasticity were considered, as a random number was subtracted from the central node and random contributions were added to the neighbors. These models allowed all unstable nodes to release their energies. However, in the present model, the unstable node releases its energy during a reconnection with four degrees of stochasticity (Equation (\ref{eqq5})).

We extracted the distribution of recorded solar flare soft X-ray peak fluxes at 1-8 \AA~ from the LMSAL Latest Events Archive (C class and above) between 2003-2017. The peak fluxes are divided by $ F_{0}=10^{-6} $ $ \textrm{W}/\textrm{m}^{2} $. Figure \ref{fig3}I shows the logarithmically-binned histogram for the LMSAL data. The observed flare peak flux distribution follows a power law over almost three orders of magnitude in peak flux. The genetic algorithm is applied to fit Equation (\ref{eq9}) and the obtained power-law index is 2.201 $\pm$ 0.002. The power-law index is also estimated using a maximum likelihood estimation (MLE) approach and the result is 2.21 $ \pm $ 0.02. As we stated in Section \ref{MLE}, MLE is a method to infer the power-law index and the lower threshold which avoids binning the data. Figure \ref{fig3}I shows also the two model PDFs. The cumulative distributions for the data, and for the two fitted models are shown in Figure \ref{fig3}II. The parameters (the power-law index,  threshold, and normalization coefficient) of these two models are the same as in panel I, and are obtained by optimization of the chi-square function and application of the MLE method on the PDF. The CDFs use Equation (\ref{eqq102}) with $ S_{2} $ equal to the maximum observed peak flux. See Section \ref{sec:style} for details of the fitting procedure.

We investigated the statistics of total released energies of the simulated flaring events. The total released energy is obtained by integrating over all liberated energies during bursts within each time step. For the case A (Figure \ref{fig4}I,II), the power-law indexes of the total released energies are 2.053 $\pm$ 0.002 and 2.07 $ \pm $ 0.03, applying the genetic algorithm and MLE approaches respectively.

We also studied the statistics of simulated peaks, as the maximum released energy among all bursts within a time step. The power-law indexes of simulated peaks for the case A are 1.843 $\pm$ 0.005 and 1.84 $\pm$ 0.02, using the genetic algorithm and MLE approaches respectively (Figure \ref{fig4}III,IV). These values are in good agreement with reports of power-law indexes for soft X-ray peak fluxes for flares, in the range 1.65-2.1 \citep{yashiro2006, aschwfree2012, mendoza2014, aschwa2015}. The obtained index for total energy from the simulation is also compatible with our analysis of the soft X-ray peak-flux observational data (Figure \ref{fig3}).

Figure \ref{fig55} shows the differential and cumulative distributions for total released energies of case C. As shown in the figure, the distribution of the simulated flaring events of the case C is not obviously a power law.

A brief summary of the obtained power-law indexes for released energies of flaring events, based on the chi-square fitting technique and MLE is presented in Table \ref{table1}.

\subsection{Statistics of Waiting Times of Flaring Events} The waiting times ($ \triangle t $) of the recorded peak fluxes from the LMSAL Latest Event Archive are shown in Figure \ref{fig3}II,IV. The waiting times are scaled by $ t_{0}, $ where the normalization factor $ t_{0} $ is the average waiting time, $ t_{0}=(t_{\textrm{max}}-t_{\textrm{min}})/N_{\rm{e}}, $ where $ N_{\rm{e}} $ is the number of recorded events. The statistics of solar flare waiting times has been subject to a wide range of studies \citep{pearce1993, biesecker1994, Wheatland1998, mike2000, Litvinenko2002}. \cite{aschw2010} studied the WTDs of solar flares observed in hard X-rays recorded by the Reuven Ramaty High Energy Solar Spectroscopic Imager (RHESSI), the Compton Gamma Ray Observatory (CGRO), and the Solar Maximum Mission (SMM). They used the framework which was investigated by \cite{Wheatland1998} and \cite{Litvinenko2002}. The WTD for a stationary (constant-rate) Poisson random process is an exponential distribution $ \lambda {\text{e}}^{-\lambda \Delta t,} $ where $ \lambda $ is the average event occurrence rate. Observed flare WTDs may be modeled as a time-dependent Poisson process, involving superposition of multiple exponential functions. Briefly, \cite{aschw2010} investigated a single $ \lambda $ (a stationary Poisson process), and a variety of models with different distributions of occurrence rate (non-stationary Poisson processes). They showed that the behavior of waiting times modeled as a Poisson process with a rate distribution which is a power law times an exponential is described by $ P(\Delta t)=\lambda_{0}(1+\lambda_{0}\Delta t)^{-\alpha}, $ where $ \lambda_{0} $ is the mean occurrence rate. They showed that this single function can fit observational flare waiting times, although $\alpha$ varies with the solar cycle ($\alpha$ = 1.4 at solar minimum, and $\alpha$ = 3.2 at solar maximum). Applying the same approach, we use an equivalent thresholded power-law function (Equation (\ref{eq9})) to study the waiting times of the observational events. The genetic algorithm is applied over the logarithmically binned histogram of the waiting times (Figure \ref{fig5}I), in order to achieve the best fit. The power-law index obtained is 1.91 $\pm$ 0.02. The estimated power-law index with the MLE approach is 1.90 $ \pm $ 0.03.

In order to construct the waiting-time distribution for the simulated flaring events, we constructed a lattice with size 104 $ \times $ 104, divided into four subsets, representing four flux tubes in four active region on the Sun's surface. The initial configurations of the subsets are independent from each other. The boundary conditions of each subset are assumed to be open and also independent. The evolutions (excitements and redistributions) of all the subsets are studied, simultaneously. Therefore, the time interval (waiting times) between two successive flaring events is determined by considering the order of flaring events within all subsets, so if the first flare occurs in subset number 1, the next flare could be triggered in the same panel or the other ones. The waiting-time distributions of the simulated flaring events are shown in the right panel of Figure \ref{fig5}. We fitted a double exponential function
\begin{eqnarray}
\label{eq16}
P(\Delta t)= a \lambda_{1} \text{e}^{-\lambda_{1}\Delta t}+ b \lambda_{2} \text{e}^{-\lambda_{2}\Delta t},
\end{eqnarray}
over the WTD. The occurrence rates obtained are $ \lambda_{1}=$ 1.16 $\pm$ 0.02, and $ \lambda_{2}=$ 0.03 $\pm$ 0.01. The corresponding non-dimensional time scales $ \tau_{i}= 1/\lambda_{i} $ for $ i=1,2 $ are about 1.4 and 71 times the average flare waiting times, respectively.

The model WTD suggests that in the model flares occur approximately as a random process with two different rates. In fact inspection of the cumulative number of events versus time shows that the mean rate of events in the model changes continuously with time. We have modeled the time rate of occurrence of events using the Bayesian blocks procedure \citep{mike2000}. This algorithm represents the data with a piecewise-constant Poisson model with rates $ \lambda_{i} $ and durations $  t_{i} $ for $ i=1,2,...,M $ (the \textquotedblleft blocks\textquotedblright). The WTD for this model is
\begin{eqnarray}
\label{eq166}
P(\Delta t)= \sum \phi_{i} \lambda_{i} \text{e}^{-\lambda_{i}\Delta t},
\end{eqnarray}
with $ \phi_{i}=\frac{\lambda_{i}t_{i}}{\sum \lambda_{i}t_{i}} $. The result is also presented in Figure \ref{fig5}II.
\subsection{Conservation of Helicity} Due to the shuffling motions of magnetic footpoints, the magnetic flux tube (and also each strand within the flux) twists. The torsion propagates along the tube axis as the driving continues. The azimuthal component of the magnetic field increases by a factor $ (1+\epsilon)^{t} $ within $ t $ iterations of excitement. In other words, the injection of helicity resulting from the photospheric motions increases the helicity of the flux tube. Thus, the magnetic helicity $ (H=\int \boldsymbol{A} \cdot \boldsymbol{B} d^{3} r) $ is not conserved during the global driving. However, $\sum A_{z}B_{z}$ remains conserved during magnetic reconnections as the magnetic vector potential is conserved during redistributions which occur within a time step. Therefore, the magnetic helicity remains conserved as the stressed strands reconnect and release energy \citep{Berger1999}.
To achieve an approximate conservation of helicity for the system superposed on a stationary state, we introduced loss at the boundaries (open boundary conditions). This loss is needed because there is very little energy dissipation in the coronal plasma with high magnetic Reynolds number.

\section{Discussion and Conclusion} \label{sec:discussion}

\begin{figure}
\centering
\includegraphics[scale=0.23]{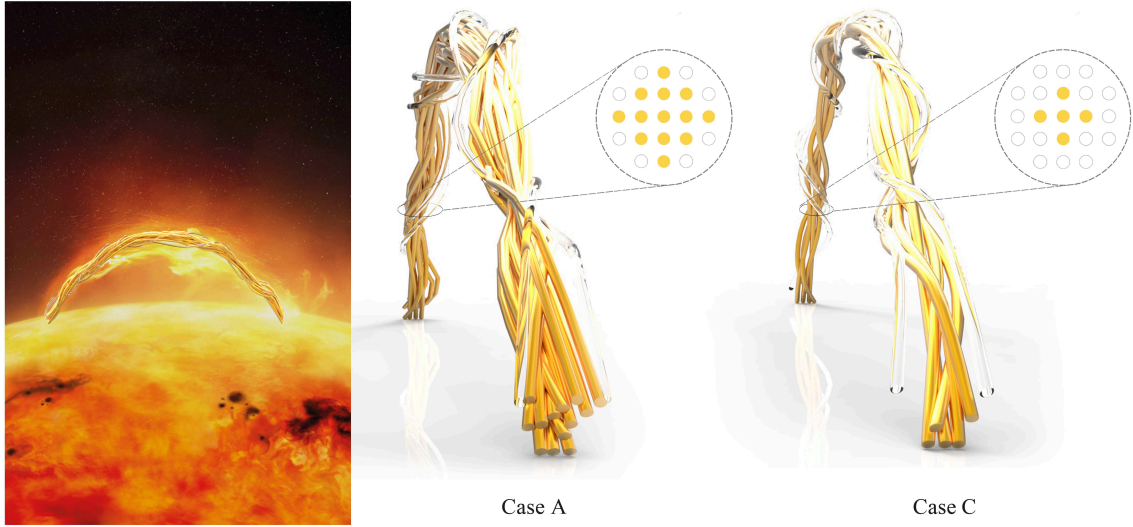}
\caption{Simple schematics of a flux tube including magnetic strands, and the dependency of the liberated energy during a redistribution on nodal values of neighboring cells in a 2D lattice (cross section of the flux tube). The strands represent the average value of the magnetic field lines (or their equivalent magnetic vector potential field) in each cell of the lattice. The evolution of the magnetic vector potential field and energy are studied in a cross section of a flux tube. Case A: Considering the lattice energy proportional to $ \sum A^{2} $, the released energy during a redistribution is a function of nodal values of 5 cells that are being redistributed, showing by yellow circles. Case C: Using
$ \sum E=\frac{1}{2} \boldsymbol{A}\cdot\boldsymbol{J} $
to calculate the lattice energy, the released energy during a redistribution depends on nodal values of all 13  neighboring cells showing in the figure, according to Equation (\ref{eqq5}).}
\label{fig1}
\end{figure}

\begin{figure}
\centering
\includegraphics[scale=0.27]{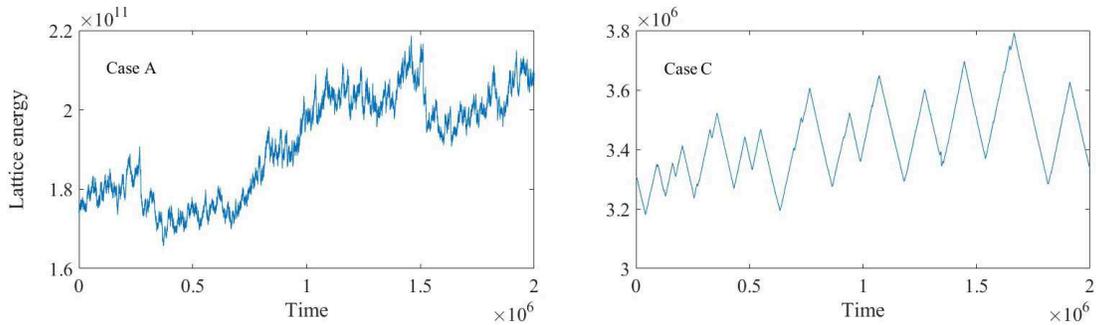}
\caption{The evolution of lattice energy. The left and right panels display the total magnetic energy of the system, which shows fluctuations on top of an approximate stationary state for cases A and C, respectively.}
\label{fig2}
\end{figure}

\begin{figure}
\centering
\includegraphics[width=18.3cm,height=8.1cm]{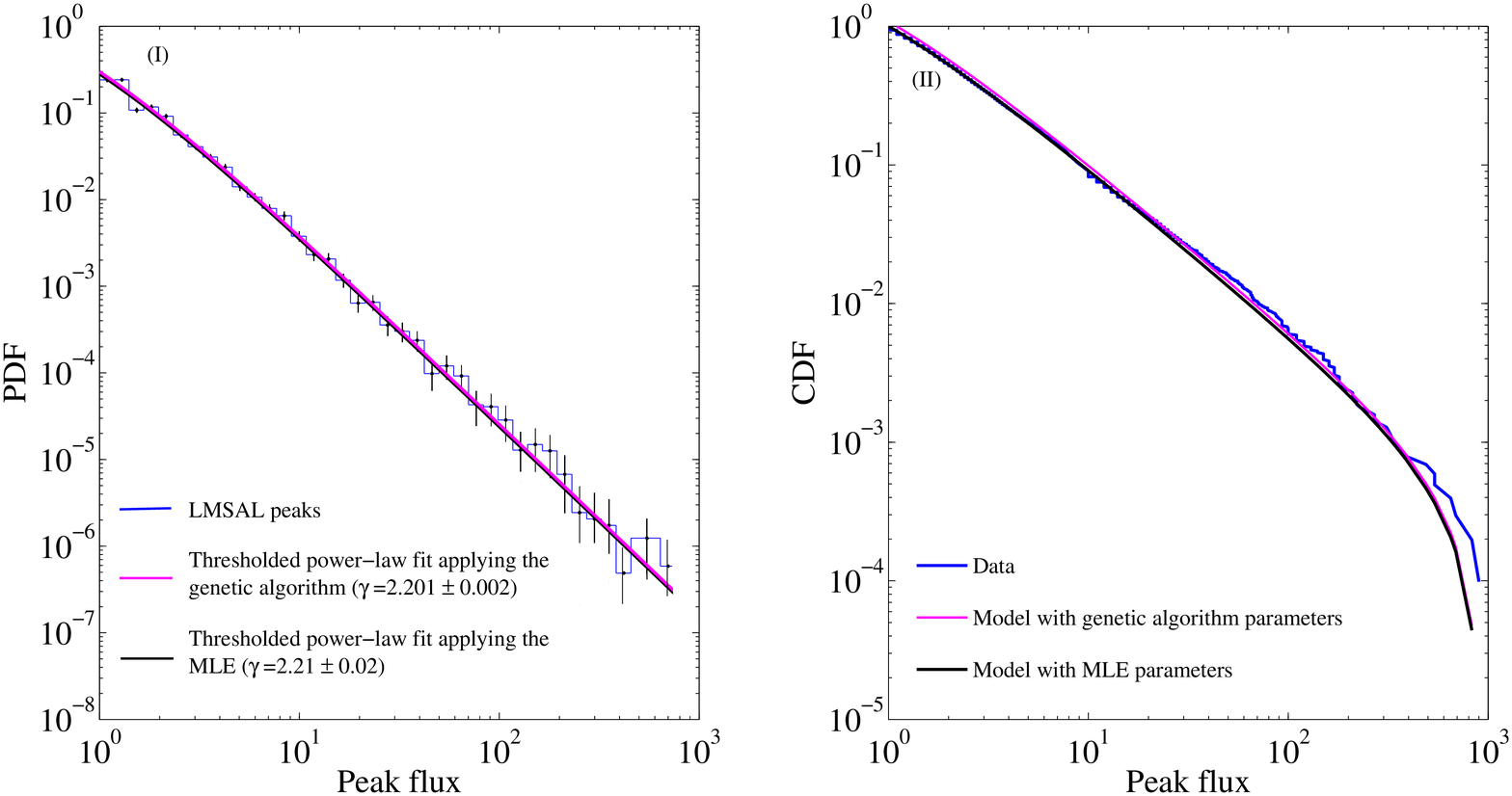}
\caption{The frequency-peak flux distribution of recorded peak fluxes from the LMSAL Latest Event Archive between 2003-2017. The peak fluxes are scaled as $ F/F_{0} $ with $ F_{0}=10^{-6} $ $\rm{W/m^{2}}$. (I) The logarithmically binned histogram of the peak fluxes and their related uncertainties, and the model PDFs. The best fits are obtained using the genetic algorithm to minimize the chi-square function of Equation (\ref{eqq10}), and using also the maximum likelihood estimation (MLE) in the Bayesian framework. The power-law indexes of the fits to the peak flux distribution obtained are 2.201 $\pm$ 0.002 and 2.21 $\pm$ 0.02 using the genetic algorithm and MLE methods, respectively. (II) The sorted peak fluxes in ascending order (blue line), together with the two models for the CDF. The parameters of these models are the same as in panel I.}
\label{fig3}
\end{figure}

\begin{figure}
\centering
\includegraphics[width=25.cm]{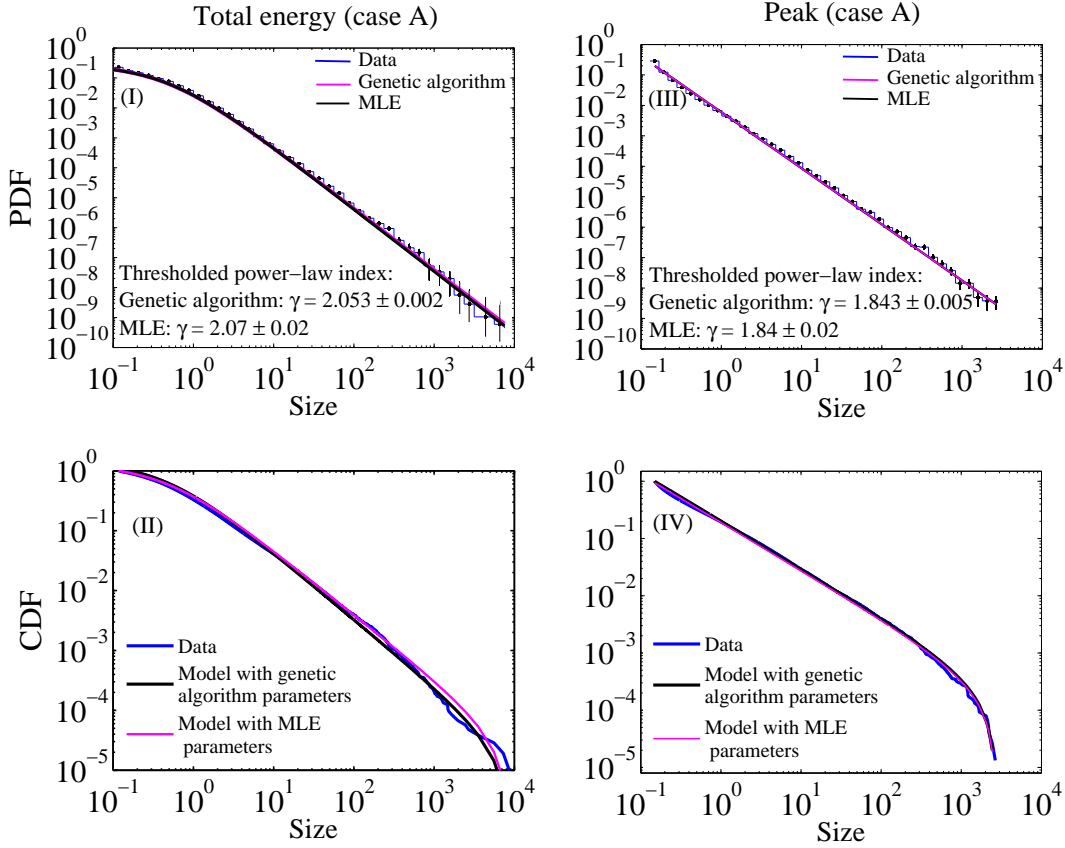}
\caption{The total released energies and peaks of the simulated flaring events of case A. The energy is considered as $ E=\sum\frac{1}{2} \boldsymbol{A}\cdot\boldsymbol{J} $. The upper panels show the logarithmically binned histograms for the data and their related uncertainties for total energy and for the peaks of the simulated events, together with the model PDFs. The lower panels show the sorted data and the model CDFs, with the same parameters for the models as in the upper panels. Applying the genetic algorithm, the power-law indexes of the total released energies and the peaks for the simulated flaring events of the case A are 2.053 $\pm$ 0.002, and 1.843 $\pm$ 0.005, respectively. Using the maximum likelihood estimation method, the power-law indexes inferred from the power law plus threshold model are 2.07 $ \pm $ 0.02 and 1.84 $\pm$ 0.02.}
\label{fig4}
\end{figure}

\begin{figure}
\centering
\includegraphics[width=18.3cm,height=8.1cm]{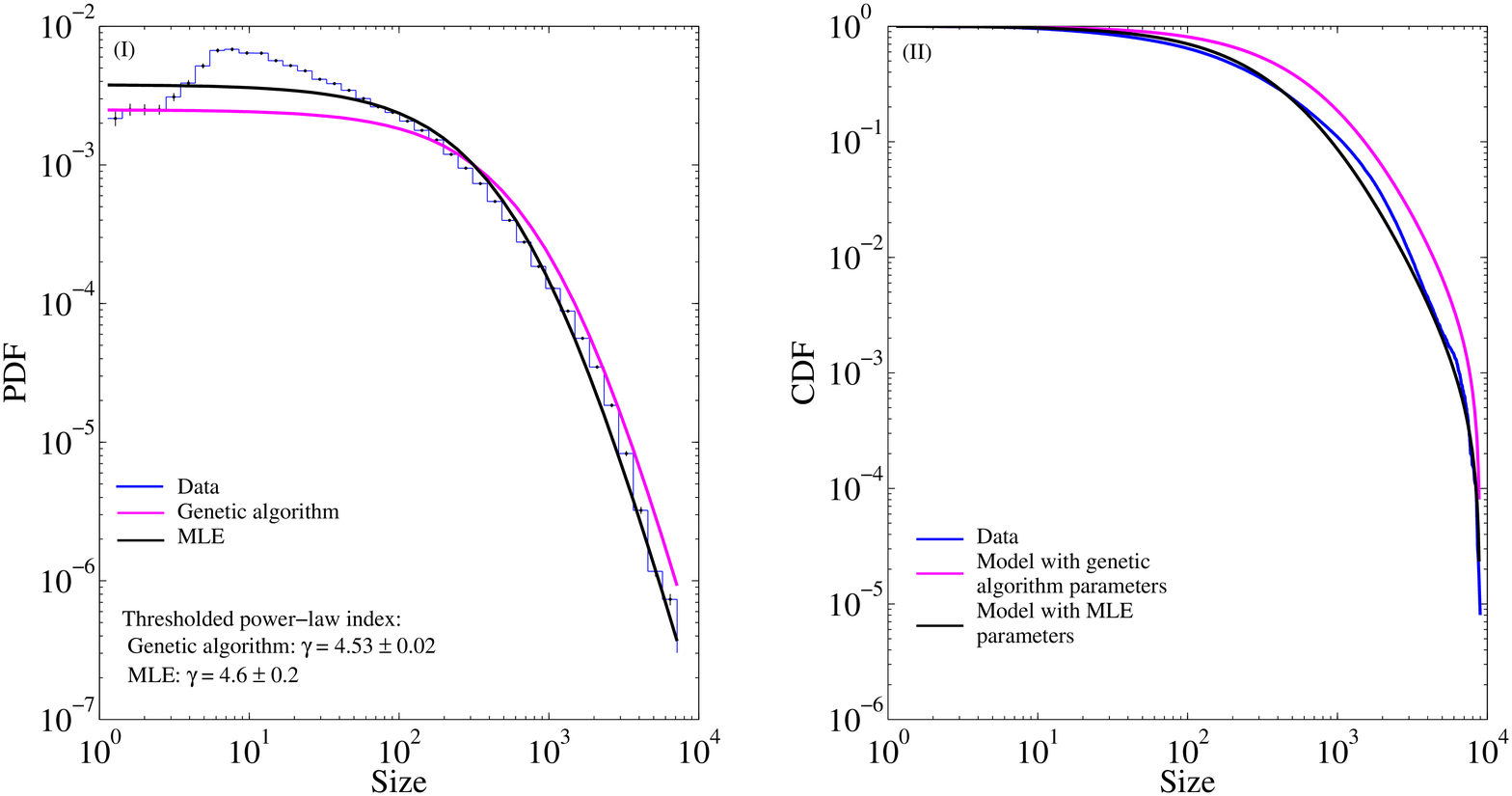}
\caption{The total released energies of the simulated flaring events of case C. The left panel shows the logarithmically binned histogram for the data and their related uncertainties for total energy of the simulated events. The energy is considered as $ E=\sum A^{2} $. The right panel shows the sorted data and the model CDF, with the same parameters for the model as in the left panel.}
\label{fig55}
\end{figure}

\begin{figure}
\centering
\includegraphics[width=18.3cm,height=8.1cm]{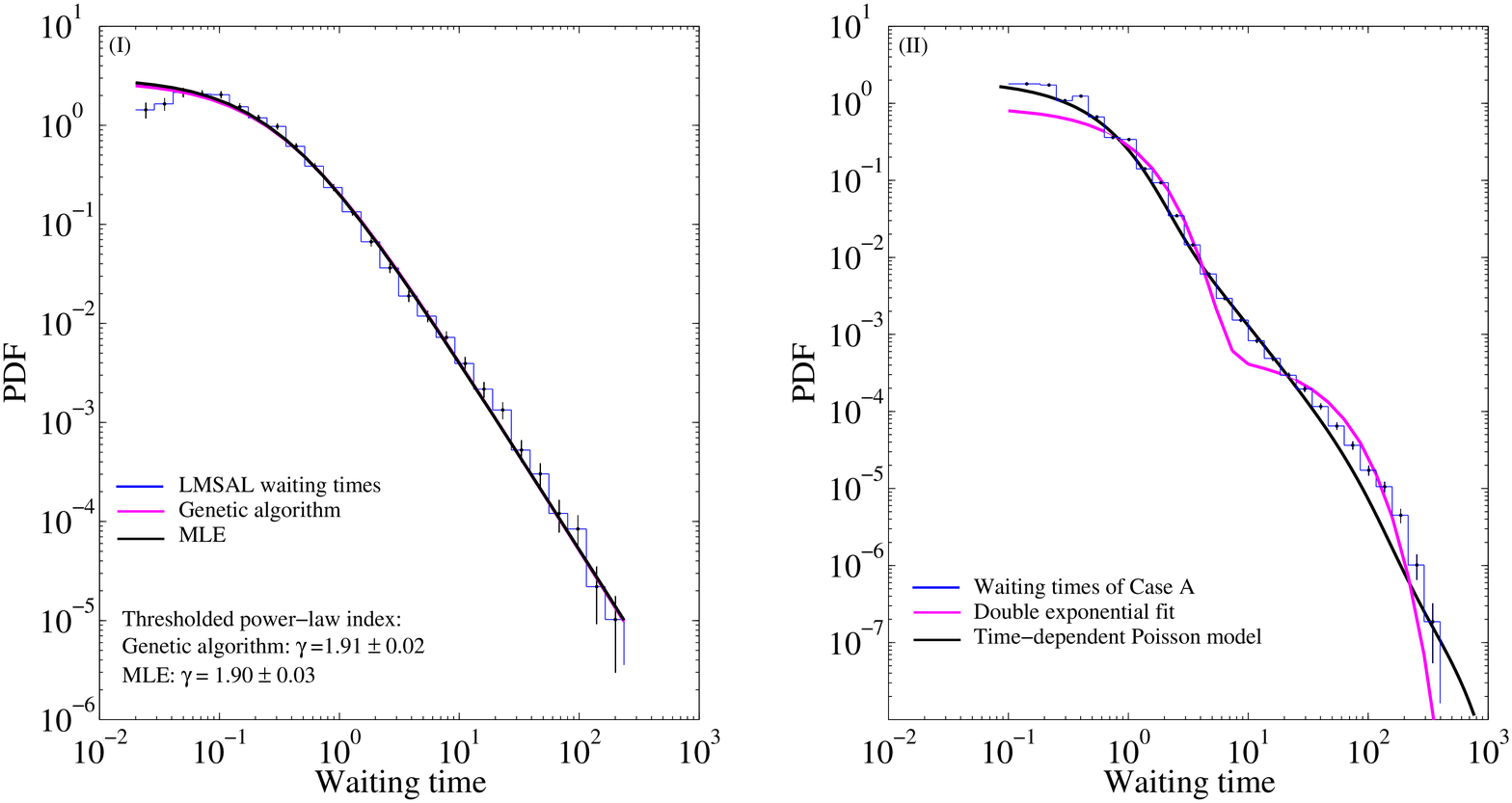}
\caption{The waiting-time distributions of both observational and simulated flaring events (case A). (I) The logarithmically binned distribution of waiting times from the LMSAL Latest Event Archive between 2003-2017. The waiting times are normalized by the average waiting time $ t_{0}=(t_{\mathrm{max}}-t_{\mathrm{min}})/N_{\mathrm{e}} $, where $ N_{\mathrm{e}} $ is the number of recorded events. The best fit is obtained using the genetic algorithm. The obtained power-law index of the fit for each method is 1.91 $ \pm $ 0.02. (II) The logarithmically binned histogram of waiting times of the simulated flaring events, for case A. The double exponential function of Equation (\ref{eq16}) is fitted to the data using the genetic algorithm (red line). The obtained occurrence rates are $ \lambda_{1}=$ 1.16 $\pm$ 0.02, and $ \lambda_{2}=$ 0.03 $\pm$ 0.01. The black line shows the WTD for a piecewise-constant Poisson model, with rates estimated from the simulation data using the Bayesian blocks algorithm.}
\label{fig5}
\end{figure}

In this study, we presented a new avalanche model for solar flares which maximizes the liberated magnetic energy, according to the principle of minimum energy. We constructed a two-dimensional lattice of magnetic vector potential instead of magnetic field \citep{LH1991, LH1993, isliker1998, takalo1999}, which could not satisfy $ \boldsymbol{\nabla} \cdot \boldsymbol{B}=0 $. The lattice represented the cross section of a flaring magnetic flux tube, with a background magnetic field along the tube axis. An open boundary condition is considered for the lattice. We applied a global driving with rate $ \epsilon $ to our model, assumed to originate from the turbulent flows and twisting of the magnetic footpoints of solar coronal loops. Within $ t $ iterations of excitement, the $z$-component of the vector potential changes proportional to $ (1+\epsilon)^{t} $. The open boundary condition (zeroing out at edges of the lattice) means that the self organized critical state mostly occurs in the central cells of the tube. This provides a near axisymmetry about the tube center. Therefore, the $z$-component of the magnetic field does not change during excitements. This implies that only the azimuthal component of the magnetic field increases by a factor of $ (1+\epsilon)^{t}. $ This stresses the magnetic field, which relaxes by liberating the stored energy through magnetic reconnections between neighboring strands. Magnetic energy is released under an evolutionary process and approaches a minimum value at a final equilibrium, as stated by the principle of minimum energy. During the redistributions/reconnections the magnetic helicity remains conserved as the magnetic vector potential is conserved. Conservation of magnetic helicity is a fundamental fact in dissipative events such as solar flares. However, the magnetic helicity is not conserved during the global excitation. The avalanches (sequence of bursts) occur as fluctuations on an approximate stationary state, in which the energy output (dissipation via reconnection and loss at the open boundary conditions) is balanced by the global driving. The magnetic energy of the system is computed using two methods: an accurate method using $ E=\sum\frac{1}{2} \boldsymbol{A}\cdot\boldsymbol{J}, $ and an approximate method using $ E=\sum A^{2}.$ The main difference between these two methods is due to the contribution of the neighboring magnetic fields (strands) in the interactions with the central magnetic field, since the number of grid point involved in the accurate method is more than twice that of the approximate method (Figure \ref{fig1}). Applying the second derivative test, it is found that the accurate method leads to the maximization of the released energy at each redistribution. Therefore, the total released energy within a time step is the maximum possible. However, through the approximate method the released energy at each local redistribution could be either minimum or maximum. Liberating an arbitrary amount of energy during a natural evolutionary process is not consistent with the principle of minimum energy, which states that the maximum possible amount of energy should be released at such process. This important fact has not been considered in the previous avalanche models.

We investigated the statistics of observed Geostationary Operational Environmental Satellite (GOES) soft X-ray flares (C class and above), using the information in the Lockheed Martin Space and Astrophysics Laboratory (LMSAL) Latest Event Archive (namely the peak flux, and the start and end times of flares). We used a genetic algorithm chi-square minimization on the logarithmically binned observational and simulated flaring events to extract the indexes for the thresholded power-law distributions. We also presented a new technique based on maximum likelihood estimation in the Bayesian framework, considering a lower threshold (without binning). For a specific simulation setup ($ \overline{Z_{c}}=1 $, $ \sigma=0.01 $), the power-law indexes for the total released energy and waiting-time distributions are 2.12 and 2.1, respectively. These results are in good agreement with the LMSAL/GOES event peak fluxes, and the waiting times between the peaks.

We conclude that the presented avalanche model, in which the system releases the maximum amount of energy at an unstable magnetic field considering the principle of minimum energy, can describe the mechanism behind solar flares. In this two-dimensional model, instead of five degrees of stochasticity, four degrees are considered in the reconnection process to avoid the complete randomness of redistributions. This has resolved the deviation of temporal correlations of previous models from the solar flare observation \citep{isliker2000, hughes2003, morales2008b, morales2008a, stru2014}. The principle of minimum energy, which we applied to our self-organized critical model by reducing one degree of stochasticity compare to the previous models can be used in other cellular automaton avalanche models (e.g. for magnetic systems, flux lines in superconductors, microfracturing process, earthquakes, physiological phenomena, etc.) to describe the corresponding phenomena.
\\
\\
\textbf{Acknowledgments:} N. Farhang acknowledges the warm hospitality during her research visit to the Sydney Institute for Astronomy. The authors gratefully acknowledge an anonymous referee for her/his useful comments and suggestions. We also acknowledge the use of data from the Lockheed Martin Space and Astrophysics Laboratory Latest Events Archive, which uses GOES data provided by the NOAA National Geophysical Data Center (NGDC).
\\
\\
\textbf{Statement of provenance:} This is an author-created, un-copyedited version of an article accepted for publication in The Astrophysical Journal. IOP Publishing Ltd is not responsible for any errors or omissions in this version of the manuscript or any version derived from it. The Version of Record is available online at DOI: ~.

%




\clearpage

%

\begin{table}
\caption{The statistical properties of observational and simulated flaring events.}
\begin{center}
\begin{tabular}{c c c}
\hline
\multicolumn{3}{c} {Flaring events \hspace{2cm} Genetic algorithm \hspace{2.cm} Maximum likelihood estimation \hspace{0.5cm}}\\
\hline
\hline
\vspace{0.9cm}
LMSAL \hspace{4cm} & \hspace{0.5cm} $2.201$ $\pm$ $0.002$ \hspace{0.5cm} &  $2.21$ $\pm$ $0.02$ \\
Case A (total energy) \hspace{4cm} & \hspace{0.5cm} $2.053$ $\pm$ $0.002$ \hspace{0.5cm} &  $2.07$ $\pm$ $0.03$  \\
Case A (peak) \hspace{4cm} & \hspace{0.5cm} $1.843$ $\pm$ $0.005$ \hspace{0.5cm} &  $1.84$ $\pm$ $0.02$  \\
\hline

\\
\end{tabular}
\end{center}
\label{table1}
\end{table}

\clearpage

\bibliography{AAS09772R2.bib}

\begin{thebibliography}{}
\expandafter\ifx\csname natexlab\endcsname\relax\def\natexlab#1{#1}\fi
\providecommand{\url}[1]{\href{#1}{#1}}
\providecommand{\dodoi}[1]{doi:~\href{http://doi.org/#1}{\nolinkurl{#1}}}
\providecommand{\doeprint}[1]{\href{http://ascl.net/#1}{\nolinkurl{http://ascl.net/#1}}}
\providecommand{\doarXiv}[1]{\href{https://arxiv.org/abs/#1}{\nolinkurl{https://arxiv.org/abs/#1}}}

\bibitem[{{Abe} \& {Suzuki}(2006)}]{abe2006}
{Abe}, S., \& {Suzuki}, N. 2006, Nonlinear Processes in Geophysics, 13, 145

\bibitem[{{Alipour} \& {Safari}(2015)}]{alipour2015}
{Alipour}, N., \& {Safari}, H. 2015, The Astrophysical Journal, 807, 175

\bibitem[{Aschwanden(2011)}]{aschwbook2011}
Aschwanden, M. 2011, Self-Organized Criticality in Astrophysics
  (Springer-Verlag Berlin Heidelberg), XIV, 416

\bibitem[{Aschwanden {et~al.}(2014)Aschwanden, B.~Crosby, Dimitropoulou,
  Georgoulis, Hergarten, Mcateer, Milovanov, Mineshige, Morales, Nishizuka,
  Pruessner, Sanchez, Sharma, Strugarek, \& Uritsky}]{Aschwanden2014}
Aschwanden, M., B.~Crosby, N., Dimitropoulou, M., {et~al.} 2014, Space Science
  Reviews, 198

\bibitem[{{Aschwanden}(2011)}]{aschw2011a}
{Aschwanden}, M.~J. 2011, Solar Physics, 274, 99

\bibitem[{Aschwanden(2014)}]{aschw2014}
Aschwanden, M.~J. 2014, The Astrophysical Journal, 782, 54

\bibitem[{Aschwanden(2015)}]{aschwa2015}
---. 2015, The Astrophysical Journal, 814, 19

\bibitem[{{Aschwanden} \& {Freeland}(2012)}]{aschwfree2012}
{Aschwanden}, M.~J., \& {Freeland}, S.~L. 2012, The Astrophysical Journal, 754,
  112

\bibitem[{{Aschwanden} \& {McTiernan}(2010)}]{aschw2010}
{Aschwanden}, M.~J., \& {McTiernan}, J.~M. 2010, The Astrophysical Journal,
  717, 683

\bibitem[{{Aschwanden} {et~al.}(2000){Aschwanden}, {Tarbell}, {Nightingale},
  {Schrijver}, {Title}, {Kankelborg}, {Martens}, \& {Warren}}]{Aschw2000b}
{Aschwanden}, M.~J., {Tarbell}, T.~D., {Nightingale}, R.~W., {et~al.} 2000, The
  Astrophysical Journal, 535, 1047

\bibitem[{{Bai}(1993)}]{bai1993}
{Bai}, T. 1993, The Astrophysical Journal, 404, 805

\bibitem[{Bak {et~al.}(1987)Bak, Tang, \& Wiesenfeld}]{bak1987}
Bak, P., Tang, C., \& Wiesenfeld, K. 1987, Physical Review Letters, 59, 381

\bibitem[{{Barab{\'a}si} \& {Bonabeau}(2003)}]{barbasi2003}
{Barab{\'a}si}, A.-L., \& {Bonabeau}, E. 2003, Scientific American, 288, 60

\bibitem[{Barpi {et~al.}(2007)Barpi, Borri-Brunetto, \& Veneri}]{barpi2007}
Barpi, F., Borri-Brunetto, M., \& Veneri, L.~D. 2007, Journal of Cold Regions
  Engineering, 21, 121

\bibitem[{{Bauke}(2007)}]{bauke2007}
{Bauke}, H. 2007, European Physical Journal B, 58, 167

\bibitem[{{Berger}(1999)}]{Berger1999}
{Berger}, M.~A. 1999, Plasma Physics and Controlled Fusion, 41, B167

\bibitem[{{Biesecker} {et~al.}(1994){Biesecker}, {Ryan}, \&
  {Fishman}}]{biesecker1994}
{Biesecker}, D.~A., {Ryan}, J.~M., \& {Fishman}, G.~J. 1994, in American
  Institute of Physics Conference Series, Vol. 294, High-Energy Solar Phenomena
  - a New Era of Spacecraft Measurements, ed. J.~{Ryan} \& W.~T. {Vestrand},
  183--186

\bibitem[{Boffetta {et~al.}(1999)Boffetta, Carbone, Giuliani, Veltri, \&
  Vulpiani}]{boffeta1999}
Boffetta, G., Carbone, V., Giuliani, P., Veltri, P., \& Vulpiani, A. 1999,
  Physical Review Letters, 83, 4662

\bibitem[{{Buchlin} {et~al.}(2003){Buchlin}, {Aletti}, {Galtier}, {Velli},
  {Einaudi}, \& {Vial}}]{buchlin2003}
{Buchlin}, E., {Aletti}, V., {Galtier}, S., {et~al.} 2003, Astronomy and
  Astrophysics, 406, 1061

\bibitem[{Cant\'{o} \& Mart\'{i}nez-G\'{o}mez(2009)}]{canto2009}
Cant\'{o}, J., \& Mart\'{i}nez-G\'{o}mez, S. C. .~E. 2009, Astronomy and
  Astrophysics, 501, 1259

\bibitem[{Charbonneau {et~al.}(2001)Charbonneau, McIntosh, Liu, \&
  Bogdan}]{charbon2001}
Charbonneau, P., McIntosh, S.~W., Liu, H.-L., \& Bogdan, T.~J. 2001, Solar
  Physics, 203, 321

\bibitem[{Clauset {et~al.}(2009)Clauset, Shalizi, \& Newman}]{clauset2009}
Clauset, A., Shalizi, C.~R., \& Newman, M. E.~J. 2009, SIAM Review, 51, 661

\bibitem[{{Craig} \& {Wheatland}(2002)}]{craig2002}
{Craig}, I.~J.~D., \& {Wheatland}, M.~S. 2002, Solar Physics, 211, 275

\bibitem[{Crosby {et~al.}(1993)Crosby, Aschwanden, \& Dennis}]{Crosby1993}
Crosby, N.~B., Aschwanden, M.~J., \& Dennis, B.~R. 1993, Solar Physics, 143,
  275

\bibitem[{Daei {et~al.}(2017)Daei, Safari, \& Dadashi}]{daei2017}
Daei, F., Safari, H., \& Dadashi, N. 2017, The Astrophysical Journal, 845, 36

\bibitem[{Emslie {et~al.}(2012)Emslie, Dennis, Shih, Chamberlin, Mewaldt,
  Moore, Share, Vourlidas, \& Welsch}]{emslie2012}
Emslie, A.~G., Dennis, B.~R., Shih, A.~Y., {et~al.} 2012, The Astrophysical
  Journal, 759, 71

\bibitem[{{Georgoulis} \& {Vlahos}(1998)}]{georg1998}
{Georgoulis}, M.~K., \& {Vlahos}, L. 1998, Astronomy and Astrophysics, 336, 721

\bibitem[{Gheibi {et~al.}(2017)Gheibi, Safari, \& Javaherian}]{gheibi2017}
Gheibi, A., Safari, H., \& Javaherian, M. 2017, The Astrophysical Journal, 847,
  115

\bibitem[{Giles {et~al.}(2011)Giles, Feng, \& Godwin}]{Giles2011}
Giles, D., Feng, H., \& Godwin, R. 2011, Communication in Statistics- Theory
  and Methods, 42

\bibitem[{{Gold} \& {Hoyle}(1960)}]{Gold1960}
{Gold}, T., \& {Hoyle}, F. 1960, Monthly Notices of the Royal Astronomical
  Society, 120, 89

\bibitem[{Goldstein {et~al.}(2004)Goldstein, Morris, \& Yen}]{Goldstein2004}
Goldstein, M.~L., Morris, S.~A., \& Yen, G.~G. 2004, The European Physical
  Journal B - Condensed Matter and Complex Systems, 41, 255

\bibitem[{Haupt \& Haupt(1998)}]{haupt1998}
Haupt, R.~L., \& Haupt, S.~E. 1998, Practical Genetic Algorithms (New York, NY,
  USA: John Wiley \& Sons, Inc.)

\bibitem[{Hughes {et~al.}(2003)Hughes, Paczuski, Dendy, Helander, \&
  McClements}]{hughes2003}
Hughes, D., Paczuski, M., Dendy, R.~O., Helander, P., \& McClements, K.~G.
  2003, Physical Review Letters, 90, 131101

\bibitem[{{Isliker} {et~al.}(1998){Isliker}, {Anastasiadis}, {Vassiliadis}, \&
  {Vlahos}}]{isliker1998}
{Isliker}, H., {Anastasiadis}, A., {Vassiliadis}, D., \& {Vlahos}, L. 1998,
  Astronomy and Astrophysics, 335, 1085

\bibitem[{{Isliker} {et~al.}(2000){Isliker}, {Anastasiadis}, \&
  {Vlahos}}]{isliker2000}
{Isliker}, H., {Anastasiadis}, A., \& {Vlahos}, L. 2000, Astronomy and
  Astrophysics, 363, 1134

\bibitem[{Johannesson {et~al.}(2006)Johannesson, G., \&
  Gudmundsson}]{johans2006}
Johannesson, G., G., B., \& Gudmundsson, E.~H. 2006, The Astrophysical Journal
  Letters, 640, L5

\bibitem[{Kramer(2017)}]{kramer2017}
Kramer, O. 2017, Genetic Algorithm Essentials, Studies in Computational
  Intelligence (Springer International Publishing)

\bibitem[{Longcope(1996)}]{longcope1996}
Longcope, D.~W. 1996, Solar Physics, 169, 91

\bibitem[{{Lu} \& {Hamilton}(1991)}]{LH1991}
{Lu}, E.~T., \& {Hamilton}, R.~J. 1991, The Astrophysical Journal, 380, L89

\bibitem[{{Lu} {et~al.}(1993){Lu}, {Hamilton}, {McTiernan}, \&
  {Bromund}}]{LH1993}
{Lu}, E.~T., {Hamilton}, R.~J., {McTiernan}, J.~M., \& {Bromund}, K.~R. 1993,
  The Astrophysical Journal, 412, 841

\bibitem[{Maehara {et~al.}(2015)Maehara, Shibayama, Notsu, Notsu, Honda,
  Nogami, \& Shibata}]{maehara2015}
Maehara, H., Shibayama, T., Notsu, Y., {et~al.} 2015, Earth, Planets and Space,
  67, 59

\bibitem[{Mandelbrot(1975)}]{mandelbrot1975}
Mandelbrot, B.~B. 1975, Journal of Fluid Mechanics, 72, 401

\bibitem[{{Mendoza} {et~al.}(2014){Mendoza}, {Kaydul}, {de Arcangelis},
  {Andrade}, \& {Herrmann}}]{mendoza2014}
{Mendoza}, M., {Kaydul}, A., {de Arcangelis}, L., {Andrade}, Jr., J.~S., \&
  {Herrmann}, H.~J. 2014, Nature Communications, 5, 5035

\bibitem[{Mitchell(1996)}]{mitchell1996}
Mitchell, M. 1996, An Introduction to Genetic Algorithms (Cambridge, MA, USA:
  MIT Press)

\bibitem[{Morales \& Charbonneau(2008{\natexlab{a}})}]{morales2008b}
Morales, L., \& Charbonneau, P. 2008{\natexlab{a}}, The Astrophysical Journal,
  682, 654

\bibitem[{Morales \& Charbonneau(2008{\natexlab{b}})}]{morales2008a}
Morales, L.~F., \& Charbonneau, P. 2008{\natexlab{b}}, Geophysical Research
  Letters, 35, n/a

\bibitem[{Newman(2005)}]{newman2005}
Newman, M. E.~J. 2005, Contemporary Physics, 46, 323

\bibitem[{Nita {et~al.}(2002)Nita, Gary, Lanzerotti, \& Thomson}]{nita2002}
Nita, G.~M., Gary, D.~E., Lanzerotti, L.~J., \& Thomson, D.~J. 2002, The
  Astrophysical Journal, 570, 423

\bibitem[{{Parker}(1988)}]{Parker1988}
{Parker}, E.~N. 1988, The Astrophysical Journal, 330, 474

\bibitem[{{Pearce} {et~al.}(1993){Pearce}, {Rowe}, \& {Yeung}}]{pearce1993}
{Pearce}, G., {Rowe}, A.~K., \& {Yeung}, J. 1993, Astrophysics and Space
  Science, 208, 99

\bibitem[{Priest \& Forbes(2000)}]{priest2000}
Priest, E., \& Forbes, T. 2000, Magnetic Reconnection: MHD Theory and
  Applications (Cambridge University Press)

\bibitem[{Robinson(1994)}]{robinson1994}
Robinson, P.~A. 1994, Physical Review E, 49, 1984

\bibitem[{{Shimizu}(1995)}]{shimizu1995}
{Shimizu}, T. 1995, Publications of the Astronomical Society of Japan, 47, 251

\bibitem[{Strugarek \& Charbonneau(2014)}]{strugarek2014}
Strugarek, A., \& Charbonneau, P. 2014, Solar Physics, 289, 4137

\bibitem[{Strugarek {et~al.}(2014)Strugarek, Charbonneau, Joseph, \&
  Pirot}]{stru2014}
Strugarek, A., Charbonneau, P., Joseph, R., \& Pirot, D. 2014, Solar Physics,
  289, 2993

\bibitem[{{Takalo} {et~al.}(1999){Takalo}, {Timonen}, {Klimas}, {Valdivia}, \&
  {Vassiliadis}}]{takalo1999}
{Takalo}, J., {Timonen}, J., {Klimas}, A.~J., {Valdivia}, J.~A., \&
  {Vassiliadis}, D. 1999, \grl, 26, 2913

\bibitem[{{Vassiliadis} {et~al.}(1998){Vassiliadis}, {Anastasiadis},
  {Georgoulis}, \& {Vlahos}}]{Vassiliadis1998}
{Vassiliadis}, D., {Anastasiadis}, A., {Georgoulis}, M., \& {Vlahos}, L. 1998,
  The Astrophysical Journal, 509, L53

\bibitem[{Wheatland(2001)}]{mike2001}
Wheatland, M. 2001, Solar Physics, 203, 87

\bibitem[{Wheatland \& Litvinenko(2002)}]{Litvinenko2002}
Wheatland, M., \& Litvinenko, Y. 2002, Solar Physics, 211, 255

\bibitem[{Wheatland(2000)}]{mike2000}
Wheatland, M.~S. 2000, The Astrophysical Journal Letters, 536, L109

\bibitem[{Wheatland \& Litvinenko(2001)}]{Litvinenko2001}
Wheatland, M.~S., \& Litvinenko, Y.~E. 2001, The Astrophysical Journal, 557,
  332

\bibitem[{Wheatland \& Sturrock(1996)}]{Wheatland1996}
Wheatland, M.~S., \& Sturrock, P.~A. 1996, The Astrophysical Journal, 471, 1044

\bibitem[{Wheatland {et~al.}(1998)Wheatland, Sturrock, \&
  McTiernan}]{Wheatland1998}
Wheatland, M.~S., Sturrock, P.~A., \& McTiernan, J.~M. 1998, The Astrophysical
  Journal, 509, 448

\bibitem[{{Yashiro} {et~al.}(2006){Yashiro}, {Akiyama}, {Gopalswamy}, \&
  {Howard}}]{yashiro2006}
{Yashiro}, S., {Akiyama}, S., {Gopalswamy}, N., \& {Howard}, R.~A. 2006, The
  Astrophysical Journal Letters, 650, L143

\bibitem[{{Zirker} \& {Cleveland}(1993)}]{zirker1993}
{Zirker}, J.~B., \& {Cleveland}, F.~M. 1993, Solar Physics, 144, 341

\end{thebibliography}



\end{document}